\documentclass[10pt]{article}
\usepackage[dvips]{graphicx}
\usepackage{fullpage}

\begin{document}
\title{Complexity of Langton's Ant}
\author{A. Gajardo \and A. Moreira \and E. Goles \\
Center for Mathematical Modeling and \\
Departamento de Ingenier\'{\i}a Matem\'atica \\ FCFM, U. de Chile
Casilla 170/3-Correo 3, Santiago, Chile}
\date{}

\maketitle

\begin{abstract}
  The virtual ant introduced by C. Langton has an interesting behavior,
  which has been studied in several contexts. Here we give a construction
  to calculate any boolean circuit with the trajectory of a single ant.
  This proves the P-hardness of the system and implies, through the
  simulation of one dimensional cellular automata and Turing machines, the
  universality of the ant and the undecidability of some problems
  associated to it. \\
  Complements and applet at {\tt http://www.dim.uchile.cl/~agajardo/langton}
\end{abstract}

\section{Introduction}

  The virtual ant is a system defined by
  C. Langton~\cite{langton,gale1,gale4} on the two-dimensional square
  lattice: each cell is in one of two states, {\em to-left\/} or
  {\em to-right\/}. The ant is an arrow between two adjacent cells.
  It moves one cell forward at each time step, in the direction 
  it is heading. That direction is changed according to the state
  of the cell where the ant arrives (it turns to the left or to the
  right); the state of the cell is changed after the ant's visit. The
  ant may be seen as a cellular automaton with von Neumann's
  neighborhood, or as the head of a  two-dimensional Turing machine.
  Interesting behavior follows: a single ant, starting with all cells
  in {\em to-left\/} state, has a more or less symmetric trajectory in the
  first 500 steps; then it goes seemingly randomly for about 10,000 steps,
  until it suddenly starts building an infinite diagonal ``highway'' 
  (a periodic motion with drift).

  The complex behavior of such a ``simple'' system motivated several
  studies, both experimental and analytical, as well as some generalizations
  and variants. The rule has been generalized to allow more than two
  states of the cell \cite{beuret,buni3,cohen2,gale2,gale3}, and to 
  consider several ants \cite{beuret,chopard}. It has also been
  adapted to different regular lattices \cite{cohen3,cohen1,wang,anahi1,anahi2} 
  and to planar finite graphs \cite{otropaper}.

  The ant has been studied as a paradigm for signal propagation in random
  media, in particular as a model of a particle in 2D Lorentz Lattice
  Gases when the particle interacts with the scatterers that occupy the
  space and affect its trajectory~\cite{cohen1}. 

\subsection{ Concepts from Complexity Theory }

  A {\bf decision problem} is one where the solution, for a given
  instance, is yes or no. It is said to be $decidable$ if there is an
  algorithm which answers the question in a finite time.

  Decidable problems are classified in {\bf complexity classes}, which 
  describe the amount of work needed to solve them. An important class is
  $P$: problems whose solution can be found in polynomial time.
  A problem to which any problem in $P$ can be
  reduced is called $P$-hard; if it also belongs to $P$, is called
  $P$-complete. Thus, to show that a problem is $P$-hard, it's enough
  to reduce a $P$-complete problem to it. Here we say that a problem
  $A$ is reduced to a problem $B$ (equivalently, that $B$ reduces $A$), 
  if there is a function $R$ computable using a logarithmic amount of space
  in the size of the input, such that $x$ is a positive instance of $A$ iff
  $R(x)$ is one of $B$.
  
  We say that a {\bf system} is $universal$ if it may simulate a universal 
  Turing Machine. This notion of universality implies, in particular,
  the existence of undecidable problems.
 
  The complexity and undecidability of problems associated to
  a dynamical system, as well as the existence of some kind of
  universality in it, are ways to measure the unpredictability of the 
  system. For definitions and results from Complexity Theory,
  we refer to \cite{Papadim}.

 \subsection{ Previous results }  \label{'previous'}

  There are very few results concerning the dynamics of the ant. The main
  one says that {\em for any initial configuration, the trajectory of
  the ant is unbounded.\/} \cite{buni1}.
  This has been generalized to the following: the set of cell that are
  visited infinitely often by the ant (for a given initial configuration) has
  no {\em corners} \cite{trou}.
  A corner of a set is a cell where at least two neighbors are not in
  the set, and these are not opposite to each other. 

  Unfortunately, it doesn't tell us anything else about the behavior of
  the ant in the long term. The experiments, however, suggest that the
  long-term behavior of the ant, although unbounded, is unbounded in a
  highly repetitive way. Specifically, it is conjectured: {\em
    For any initial configuration with finite support, the ant eventually
    starts building the periodic highway, in some unobstructed direction.\/
   }
  (Here, a configuration is said to have finite support if all but a
  finite number of cells are in the same state).
  If the conjecture is true, then any problem associated to the
  ant, whose input is an initial configuration with finite support,
  is decidable, for in that case, it suffices to iterate on the
  configuration until the highway appears. The question may be answered
  at that point, since the future dynamics is easily predicted.

 \subsection{ Results }

  We will present a construction that will allow the representation of any
  boolean circuit as a finite configuration in the lattice. The input
  variables are represented with the states of certain cells,
  the calculation of the circuit is performed by the trajectory of
  a single ant, and the result is written, again, as the states of certain cells.
  The first consequence of this construction is a lower bound for
  the complexity of the system: 
  \begin{quote}
   {\em There exists a problem (``does the ant ever visit this given cell?'')
   which reduces a $P$-complete problem (the calculation of a boolean 
   circuit) and therefore is $P$-hard.\/}
  \end{quote}

  On the other hand, the construction allows us to simulate any
  linear cellular automaton for configurations where only
  a finite number of cells change their states at each time. 
  The consequences are two:

  \begin{quote}
  {\em The system is capable of universal computation, since it is
  able to simulate the dynamics of a universal Turing machine.\/}
  \end{quote}
  \begin{quote}
  {\em There are undecidable problems associated to the behavior of
   the ant.\/}
  \end{quote}

\section{Construction of circuits}

\begin{figure}[p!]
  \begin{center}
  \includegraphics[width=9cm]{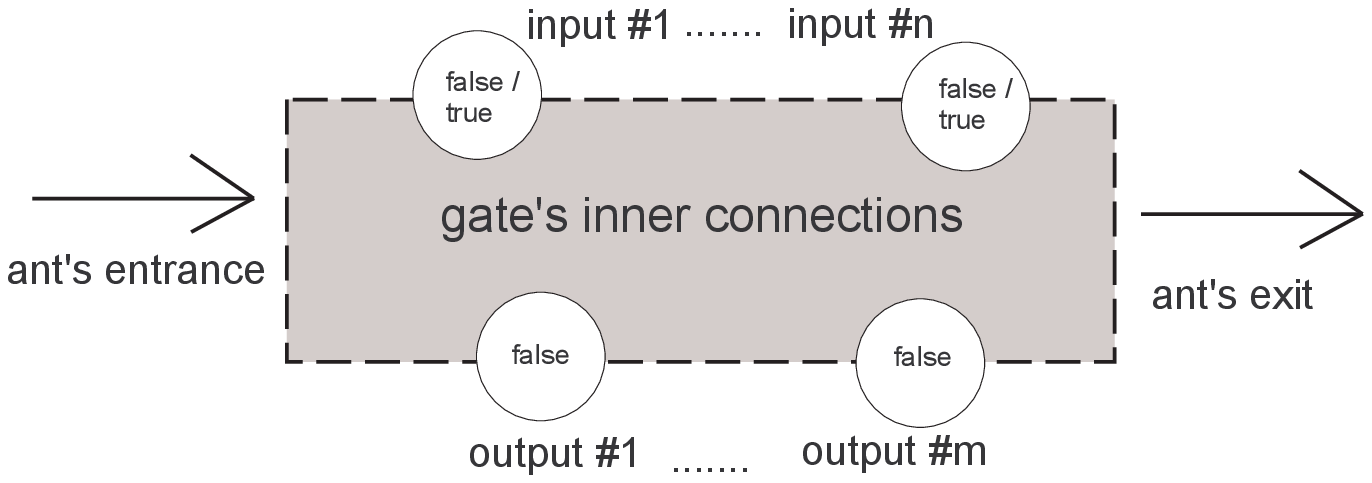}
  \end{center}
\caption{}
\label{fig:gate}
\end{figure}

  Given any boolean circuit, we'll show how to build a configuration,
  where the input variables are represented by states at certain
  locations, and the result is calculated by the trajectory of a single ant
  (and is written in a predetermined location). 
  For technical reasons, each of our one-bit input and output registers
  consist of two horizontally adjacent cells rather than one; this makes
  it possible for the ant to visit a register without visiting any other
  nearby cells in that row, and will be seen later to facilitate the re-use
  of the output register of one logic gate as an input register to a later
  gate. We construct logical gates
  with the form described in Figure \ref{fig:gate}.

  At the top of the gate, we have some cells whose states represent the
  input. At the bottom, some cells represent the output; at the beginning,
  all output cells are in the {\em to-left\/} state which will represent
  the logical value false. The ant enters the gate from the left,
  follows some path inside of the gate, and exits the gate heading to
  the right. While being in the gate, the ant visits the input cells,
  and visits (and switches) the correct output cells, according to the
  function which the gate represents. The changes are done {\em from
  inside\/}, thus allowing the output cell to be used as the input cell
  for another gate.

\begin{figure}[p!]
  \begin{center}
  \includegraphics[width=5cm]{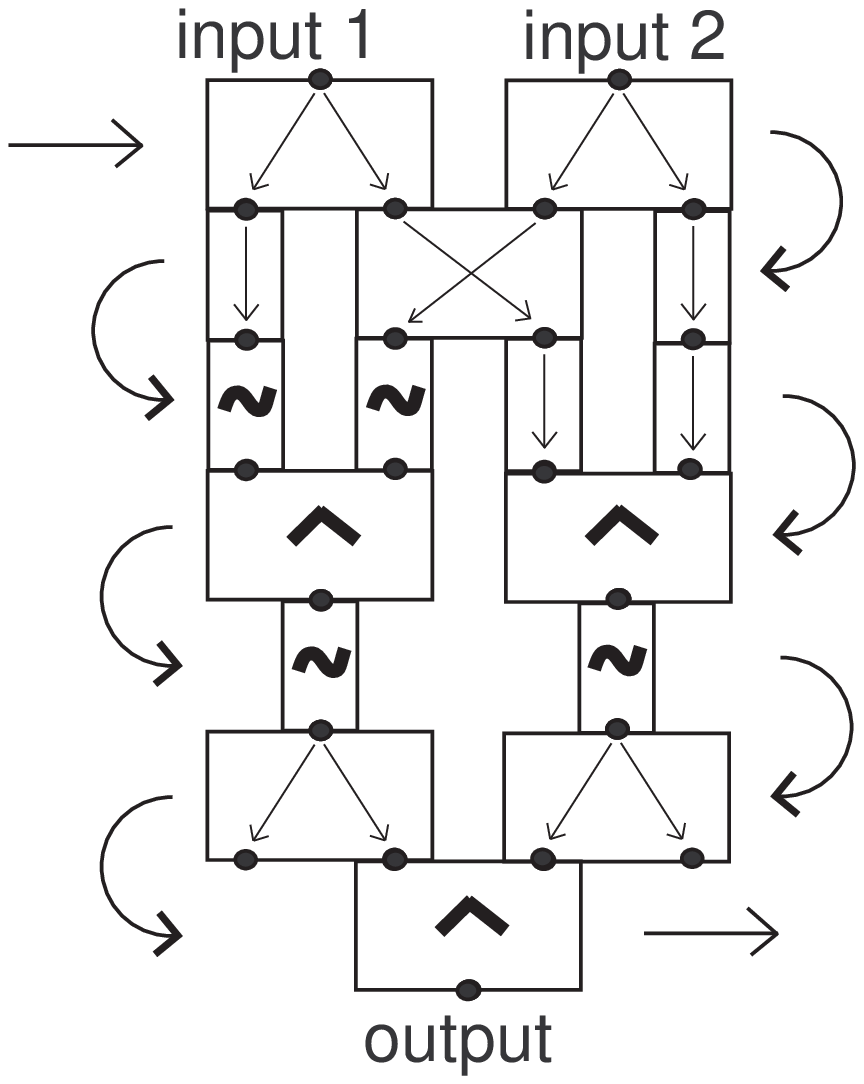}
  \end{center}
\caption{}
\label{fig:circ}
\end{figure}

\begin{figure}[p!]
  \begin{center}
  \includegraphics[width=3cm]{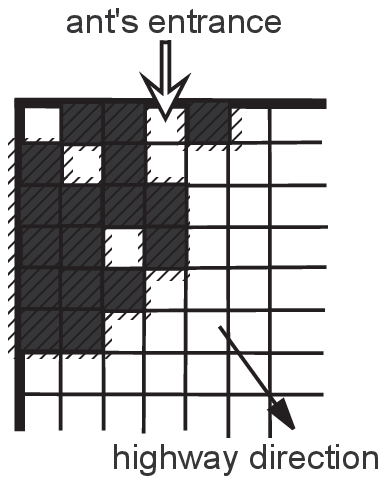}
  \end{center}
\caption{}
\label{fig:high}
\end{figure}

  To compute a boolean circuit we just put the input variables in some 
  cells at the top of the configuration (see Figure \ref{fig:circ}), and
  for the consecutive stages of evaluation we put consecutive rows of
  logical gates. The ant will go through every row, starting with the
  upper one. After going through the last row, the state of the last
  output cell will contain the evaluation of the circuit for the
  given input (to change from one row to the next we must construct
  right-to-left versions of the gates, thus allowing the ant to compute
  alternate rows in different directions). After that, we assure that
  the ant will never return to this cell by leading it to a ``highway
  seed'' (see Figure \ref{fig:high}).

  To write a boolean circuit it is enough to have the NOT and the AND
  functions. To construct the circuit we also use gates that allow us to
  duplicate, cross and copy variables.

\begin{figure}[p!]
\begin{center}
\begin{tabular}{cp{1.1cm}cp{1.1cm}c}
  Copy & & NOT & & AND \\
  \includegraphics[width=2.2cm]{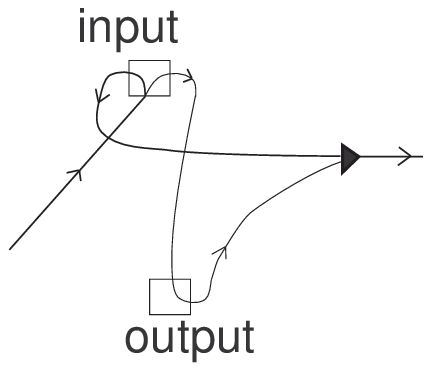} & &
  \includegraphics[width=2.3cm]{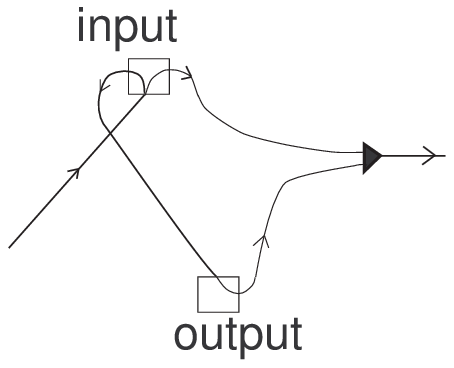} & &
  \includegraphics[width=3.2cm]{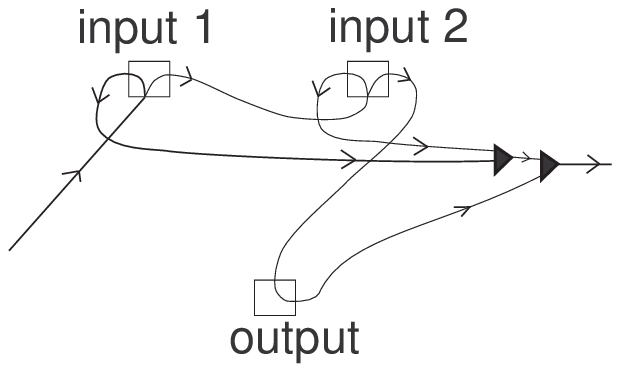}\\
& & & & \\
\end{tabular}
\begin{tabular}{cp{1.1cm}c}
  Duplicate & & Cross \\
  \includegraphics[width=2.8cm]{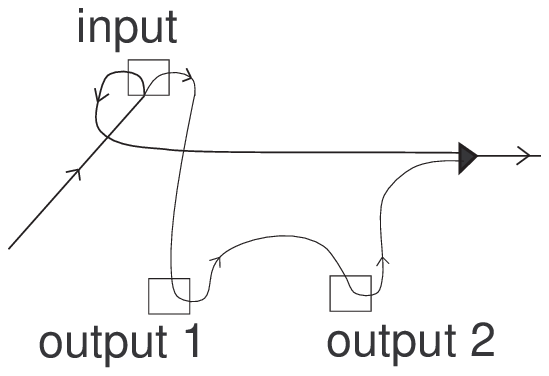}& &
  \includegraphics[width=2.9cm]{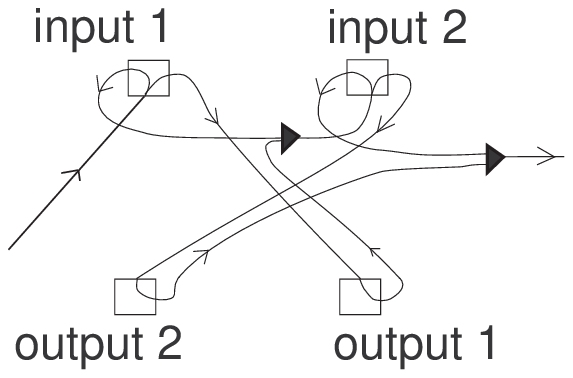}\\
\end{tabular}
  \end{center}
\caption{}   
\label{fig:not}
\end{figure}

  The way to design the gates is represented in Figure \ref{fig:not}. Let
  us describe how the NOT gate works. The ant visits the input cell, and
  depending on its state the ant will follow different paths. If the input
  cell is in the {\em to-left} state (logical value false), the ant goes to
  the output cell and  changes its state from false to true, and then 
  exits. If the input cell is in the {\em to-right} state (logical value 
  true), the ant goes directly to the exit. This is the general scheme 
  of the gates: Some of the allowed trajectories pass through the output
  register(s) and some do not, before they rejoin. The ant chooses its
  path through the gate in accordance with the input registers, either
  passes through the output register(s) (changing the states) or not,
  and then exits the gate.
  
\begin{figure}[p!]
  \begin{center}
  \includegraphics[width=4in]{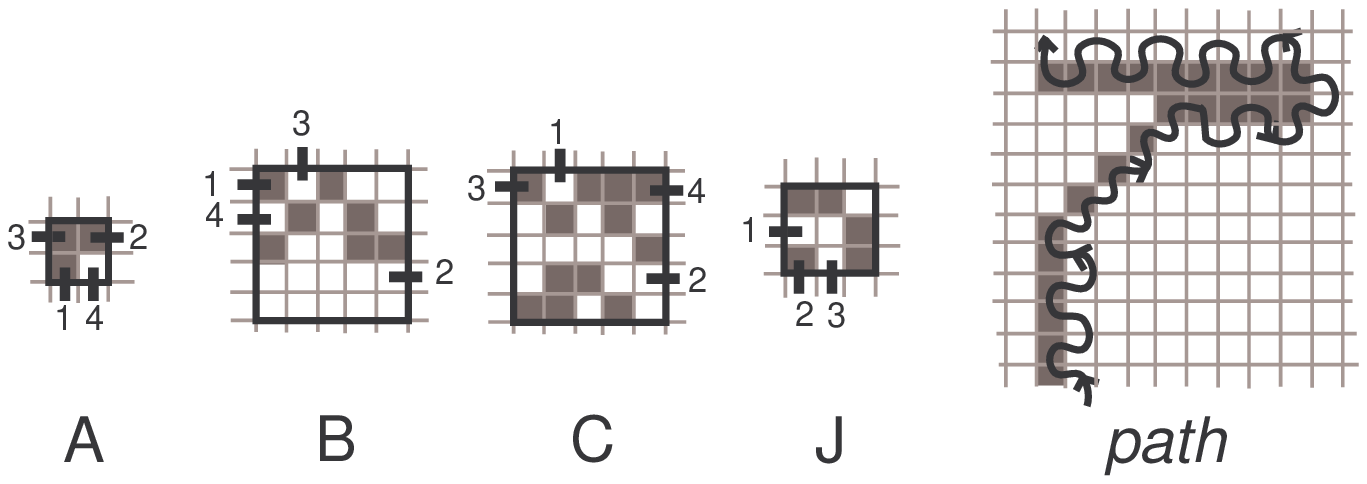}
  \end{center}
  \caption{}
  \label{fig:dev}
\end{figure}

  Now, let us see how the gates can be embedded in the lattice. 
  Figure \ref{fig:dev} shows the configurations for three kinds of
  crossings (A, B, C), the junction (J), and the paths.
  
  \begin{itemize}
   \item {\bf A:} if the ant first enters at 1, it exits at 2. If
     afterwards it enters at 3, it exits at 4 (see Figure \ref{fig:dev}).
  \item {\bf B:} if the ant first enters at 1, it exits at 2. If
     afterwards it enters at 3, it exits at 4. But if it enters first at 3,
     it also exits at 4.
  \item {\bf C:} it works as {\bf B}, but with different relative positions 
     of the input cells.
  \item {\bf J:} if the ant enters by 1 or by 2, it exits by 3.
  \item {\bf path:} is the configuration that forces the ant to follow a
     determined trajectory, allowing it to visit the input and the output
     cells in the required way. A double line with a free extremity may be
     used as a bent path, since the ant will use both sides, turning
     around the extrem.
  \end{itemize}

\begin{figure}[p!]
  \begin{center}
   \begin{tabular}{cc}
     AND & NOT \\
  \includegraphics[width=7.5cm]{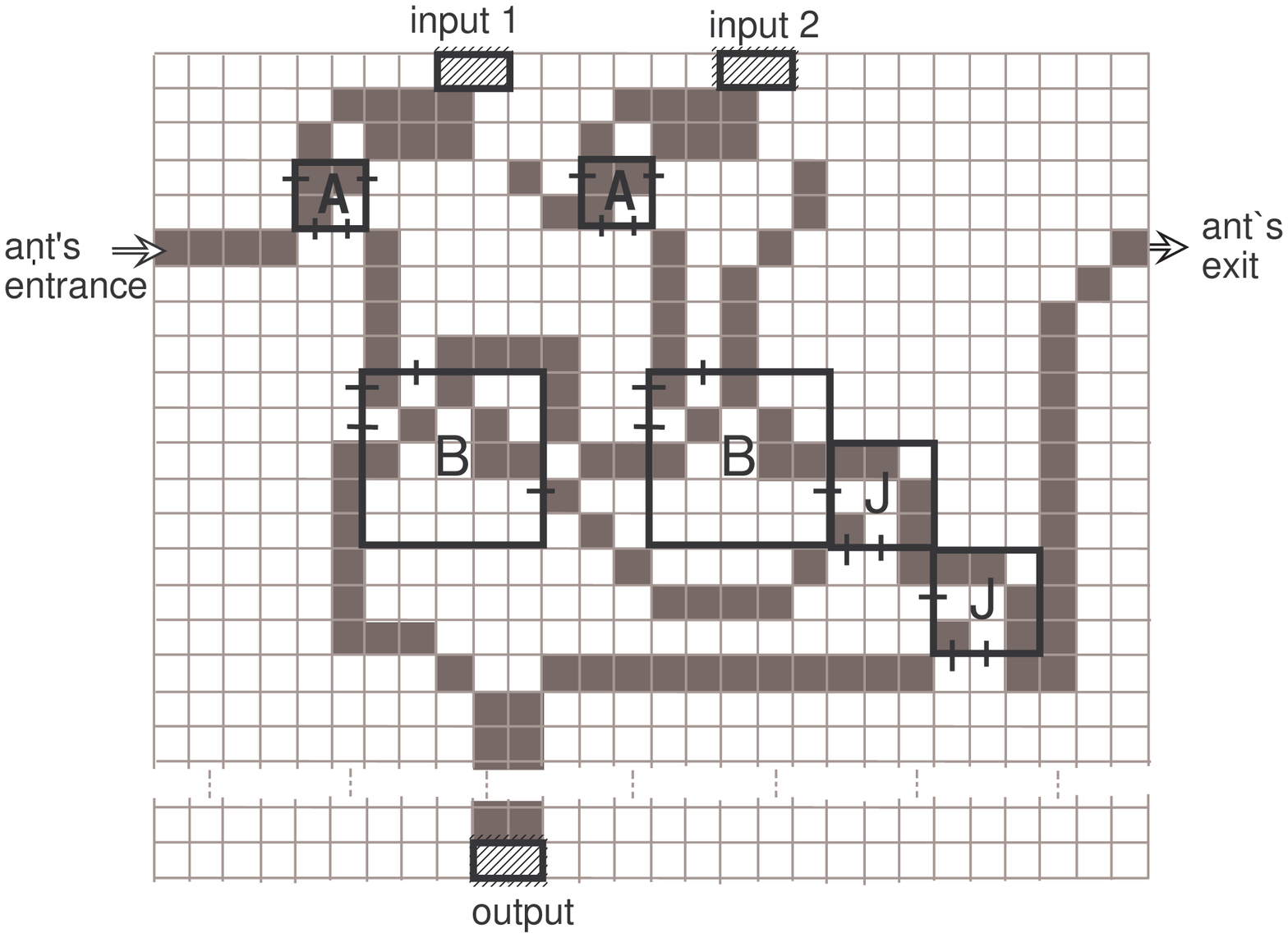} &
  \includegraphics[width=5.5cm]{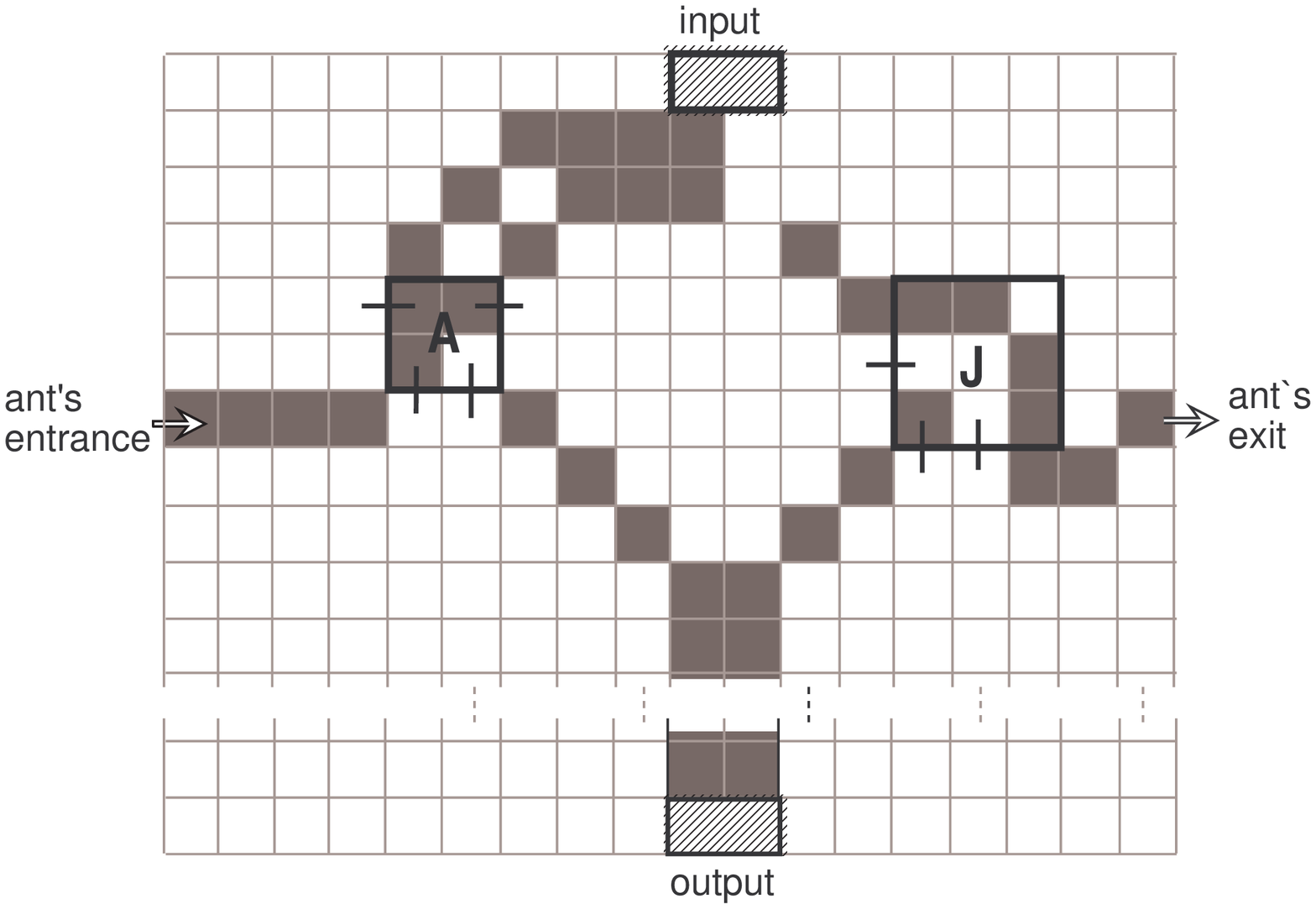} \\
   \end{tabular}
  \end{center}
  \caption{}
  \label{fig:grid1}
\end{figure}

\begin{figure}[p!]
  \begin{center}
    \begin{tabular}{c}
       Cross \\
       \includegraphics[width=9cm]{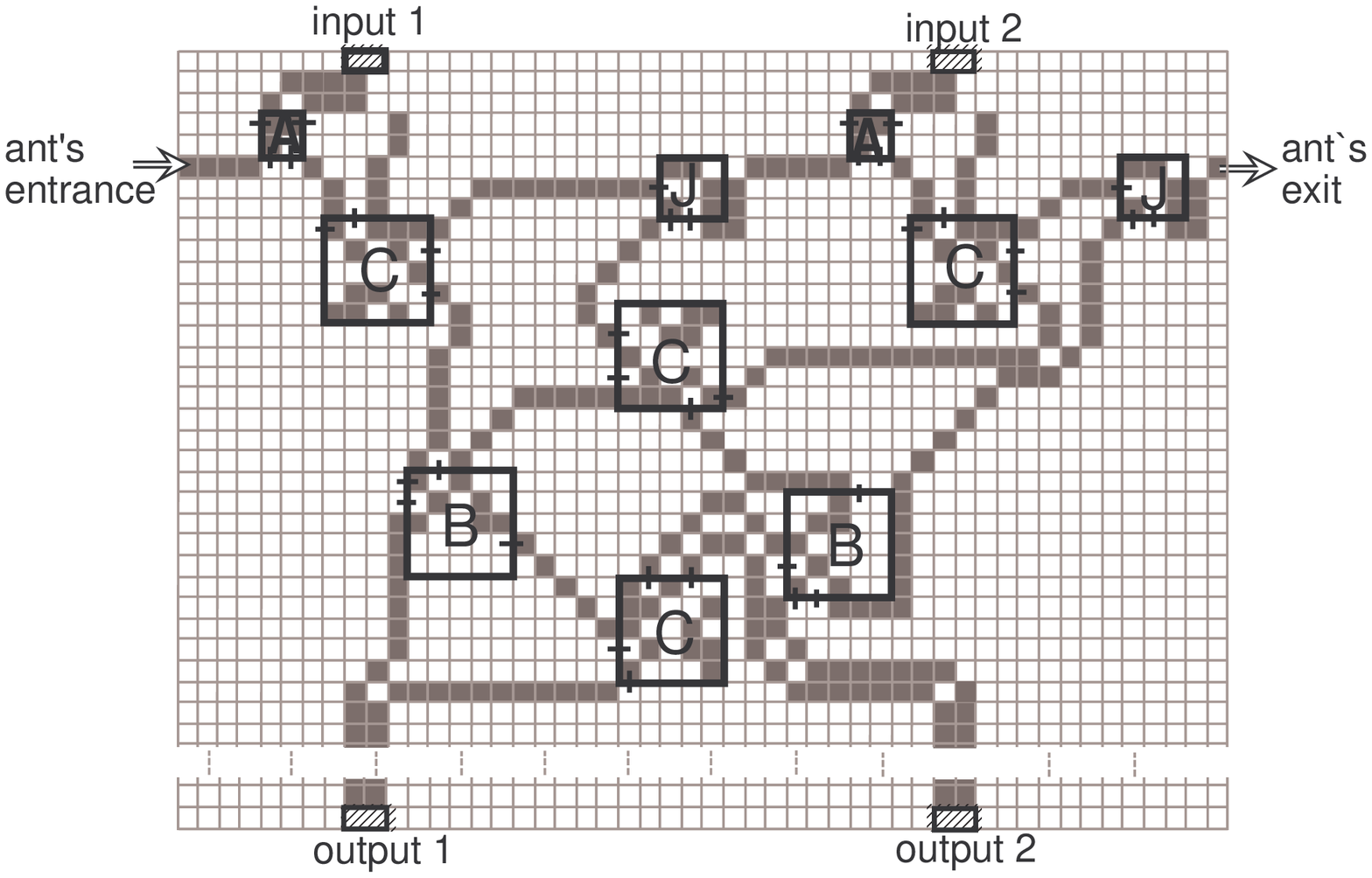}\\
    \end{tabular}
  \end{center}
  \caption{}
  \label{fig:grid2}
\end{figure}

\begin{figure}[p!]
  \begin{center}
   \begin{tabular}{cc}
    Copy & Duplicate \\
  \includegraphics[width=6.5cm]{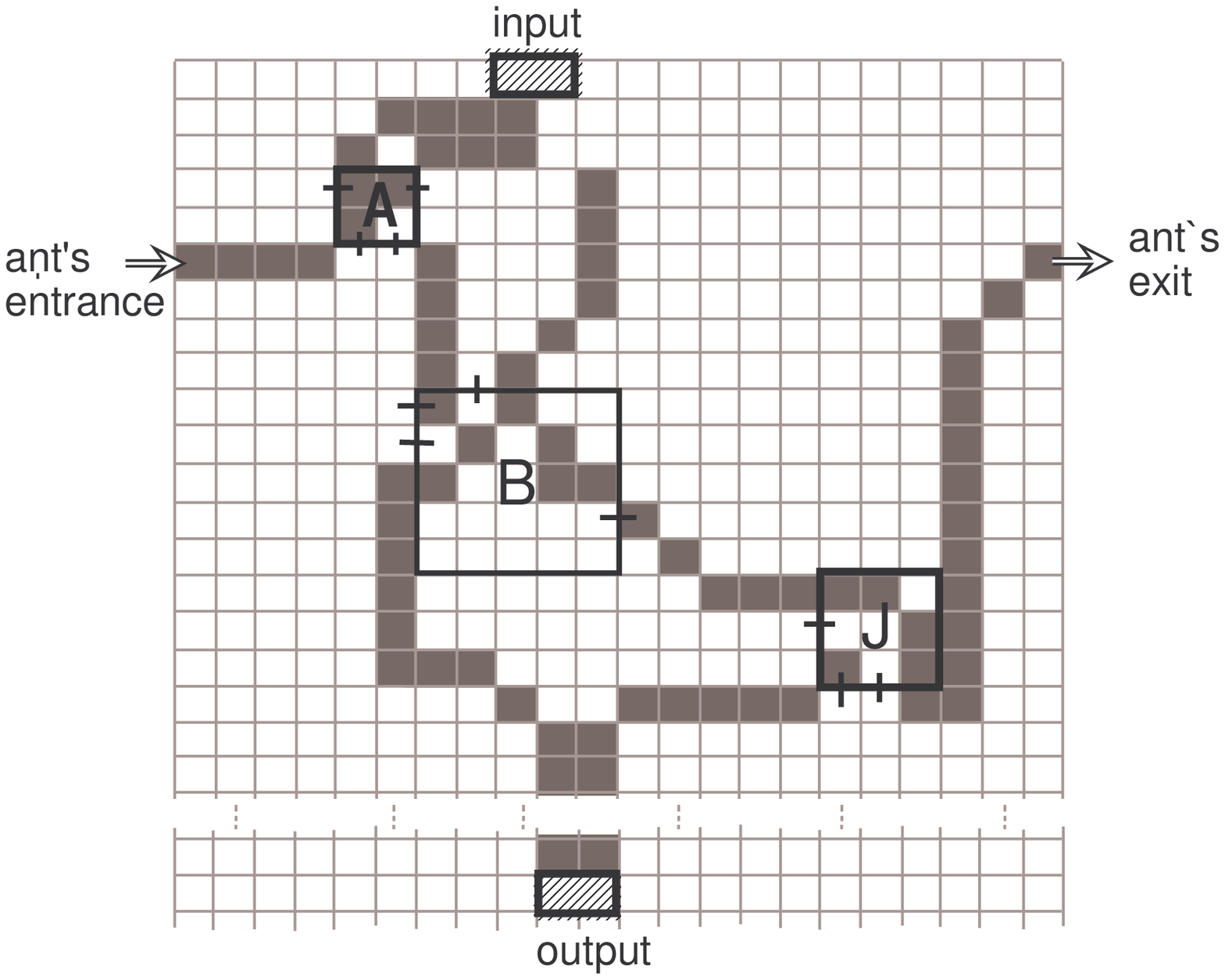} &
  \includegraphics[width=6.5cm]{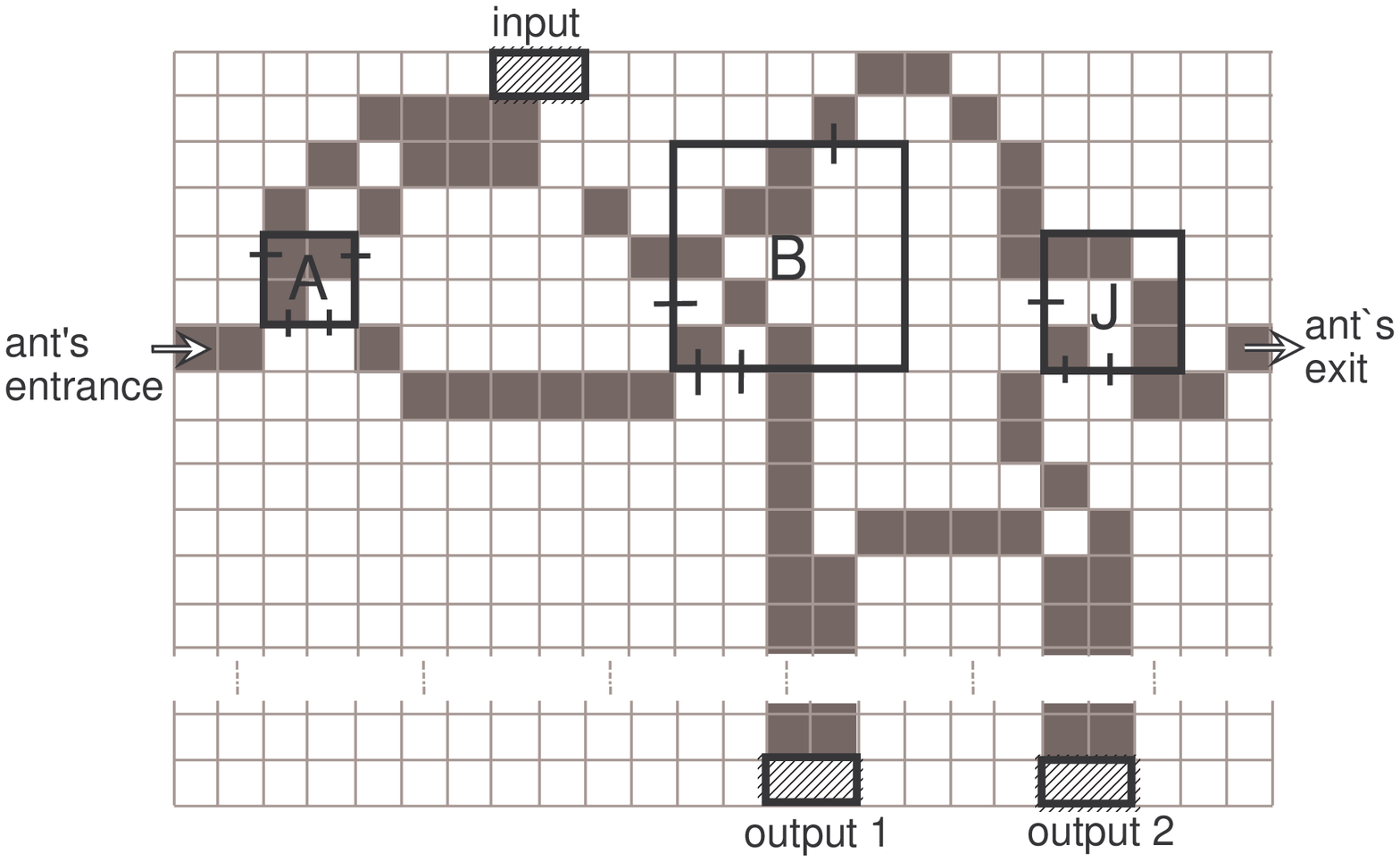} \\
 \end{tabular}
 \end{center}
  \caption{}
  \label{fig:grid3}
\end{figure}

  Figures~\ref{fig:grid1}, \ref{fig:grid2} and \ref{fig:grid3} show the
  final version of the logical gates.

\section{Discussion}

  The following problem is known to be $P$-complete \cite{Sipser}: 

  {\bf (B)} Given a boolean circuit (BC) and a truth assignment. Does the
  truth assignment satisfy (BC)?

  In the previous section we established a function that transforms an
  instance of {\bf (B)} into an instance of the following problem:

  {\bf (P)} Given a finite initial configuration of $\mathbf{Z}^2$, a given
  initial position of the ant and a cell $\alpha$. Does the ant ever visit
  $\alpha$? 

  Since an instance of {\bf (B)} is positive if and only if its image in
  {\bf (P)} is positive, the transformation is a reduction from {\bf (B)}
  to {\bf (P)}. This reduction is polynomial: the number of
  rows in the gate is bounded by two times the height $(H)$ of the BC
  plus the number of crossings, i.e., $2(H+W^2H)$, where $W$ is the
  width of the BC. The number of gates in each row is bounded by $W$. That 
  implies that the number of {\em to-right\/} cells ($S$) necessary to simulate
  an $H\times W$ BC is bounded by $2SWH(W^2+1)=o(W^3H)$. 
%
  The algorithm that defines the simulating configuration in $\mathbf{Z}^2$ needs
  only logarithmic space; all it has to do is to read and translate the
  boolean circuit. For this purpose, it has to memorize numbers such
  as the position of the symbol that is being translated and the
  current height of the circuit; these numbers are bounded by the
  size of the input, and can be recorded in polynomial space. The
  output of this ``drawing'' algorithm is the list of coordinates
  of cells in the {\em to-right} state, and is polynomial in the length
  of the input. This is all we need to legitimate the reduction,
  and we conclude that {\bf (P)} is $P$-hard.

  In a cellular automata (CA), a {\em quiescent state} is defined by the
  following property: if a cell and all its neighbors are in the quiescent
  state, the cell remains in it at the next iteration. Hence, all the dynamics
  of the system takes place at the cells in non-quiescent states and their 
  neighbors. An initial configuration with a finite support (number of
  non-quiescent states) will keep this property through the iterations 
  of the CA.

\begin{figure}[p!]
\begin{center}
  \includegraphics[width=13cm]{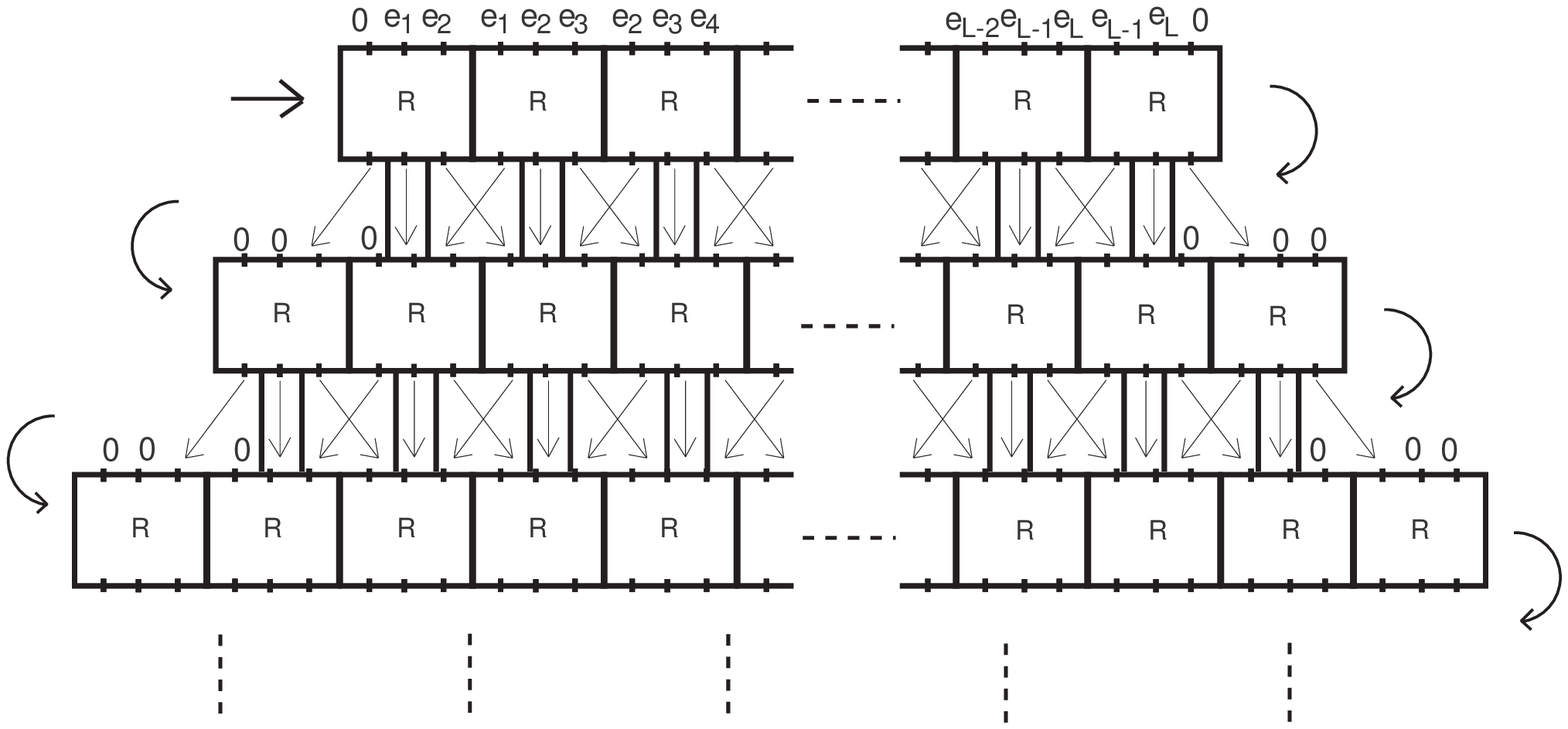}
  \end{center}
\caption{}
\label{fig:CA}
\end{figure}

  Remember that the CA transition rule can be calculated with a multi-output
  finite boolean circuit. So, for a given linear CA with quiescent state,
  we can define an initial configuration on the lattice 
  consisting of infinitely many copies of this circuit, arranged in an
  infinite trapezoidal array with top row of length $L$, as shown in
  Figure~\ref{fig:CA}.
  Any initial configuration of the CA whose support has width less than 
  $L$ can be written as the input of the first row, and the ant 
  simulates the CA.
  For widths bigger than $L$, just put the initial configuration in a
  lower row, and let the ant start running from the appropriate cell.

  It follows that the undecidability of some CA problems is inherited by
  the ant system. For instance, the problem of knowing whether a given
  (finite) word $v$ will ever appear in the evolution of a given linear CA,
  for a given initial configuration with (finite) support $u$, is reduced
  to the problem of deciding whether a given (finite) block ever appears
  in the evolution of the ant, for a given (infinite) initial configuration
  of the lattice. Since a Turing machine can be simulated by a linear CA
  with quiescent state, the ant is also universal.

\section{ Conclusions }

  The construction given here leads us to the following results about the
  dynamics of a single Langton's ant:
  \begin{itemize}
   \item
    The system is $P$-hard, in the sense that it admits a $P$-hard problem.
    The $P$-hard problem that was shown is associated with initial 
    configurations with finite support.
   \item
    The system is capable of universal computation. In spi\-te of being
    a ra\-ther weak no\-tion of u\-ni\-ver\-sa\-li\-ty 
    (which re\-qui\-res an in\-fi\-ni\-te --but finitely described-- configuration),
    it shows that the dynamics of the system is highly unpredictable.
   \item
    A direct consequence of the previous point is the existence of 
    undecidable problems. We notice that this result refers to 
    problems associated with initial configurations with infinite
    support. 
  \end{itemize}
   
  There are some open issues related to these results. One of them relates
  to the complexity; the existence of universal computation suggest that
  P-hardness is far from being a tight lower bound, and it would be 
  interesting to reduce some $NP$-complete problem to a problem of our system.

  On the other hand, the decidability of problems whose input is a
  configuration with finite support remains an open question. A positive
  answer would be given if the conjecture stated in \ref{'previous'} is
  found to be true.

\section*{Captions of figures}

{\bf Figure 1:} A sketch of a gate. The ant computes the logical gate and
changes the states of the output cells. At the beginning the output
cells are in the {\em to-left\/} state, representing the logical value
{\em false\/}.

{\bf Figure 2:} The XOR function, built upon AND, NOT, Cross, Copy and Duplicate 
gates. The ant computes the logical circuit ($\sim (i_1
\wedge i_2) \wedge \sim(\sim i_1 \wedge \sim i_2)$) row by row.  The 
circuit is satisfied iff the ant visits the output cell, for the 
given input.

{\bf Figure 3:} Highway seed. After entering at the indicated point, the ant
 never crosses the bold lines again, and starts building a highway
 in the direction of the diagonal arrow. White stands for
 the {\em to-left} state, black for {\em to-right}.

{\bf Figure 4:} A simplified scheme of the gates. The ant follows one of the
paths in each gate.

{\bf Figure 5:} Three kinds of crossings, a junction and a path.

{\bf Figure 6:} AND and NOT gates.

{\bf Figure 7:} Cross gate.
             
{\bf Figure 8:} Copy and Duplicate gates. The location of the output 
in the Copy gate can be changed in the horizontal axis, allowing us to 
fit the positions of the output variables of a row into the inputs of 
the following row.

{\bf Figure 9:} The ant simulates each iteration of the CA in a row of
gates, crosses the repetitions of the outputs (preparing the next
input) and goes to the next row. $R$ stands for the circuit that
calculates the rule.

\newpage 

\end{document}